\documentclass[12pt]{article}
\input epsf
\usepackage{graphicx}
\usepackage{bm}% bold math
\usepackage{latexsym}

\newcommand{\bea}{\begin{eqnarray}}
\newcommand{\eea}{\end{eqnarray}}
\newcommand{\be}{\begin{equation}}
\newcommand{\ee}{\end{equation}}
\newcommand{\pa}{\partial}
\newcommand{\nn}{\nonumber \\}

\newcommand{\tr}{\mbox{tr}}
\newcommand{\hphi}{\hat{\phi}}
\newcommand{\hH}{\hat{H}}

\def\href#1#2{#2}

\begin{document}

\begin{titlepage}
\vspace*{20mm}

\begin{center}
{\bf \large Matrix Model For Polyakov Loops,}\\
\vspace*{2mm}
{\bf \large String Field Theory In The Temporal Gauge, And}\\
\vspace*{2mm}
{\bf \large Winding String Condensation In Anti-de Sitter Space}\\
\vspace*{25mm}
{Furuuchi \ Kazuyuki}\\
\vspace*{3mm}
{\sl Harish-Chandra Research Institute}\\
{\sl Chhatnag Road, Jhusi, Allahabad 211 019, India}\\
{\tt furuuchi@mri.ernet.in}
\vspace*{4mm}
\begin{abstract}
It has recently been shown 
that the low energy dynamics
of the large $N$
gauge theory on $S^3$ at finite temperature
reduces to a one-matrix model,
where the matrix is 
given by the
holonomy of the gauge field 
around the Euclidean time direction
compactified on a circle.
On the other hand,
there is a prescription
for constructing a closed string field theory
in the temporal gauge
from a given one-matrix model 
via loop equations.
I identify the closed string field theory
in the temporal gauge
constructed from the above
matrix model
as 
effective closed string field theory
that describes the
propagations of closed
strings in the radial and Euclidean
time directions in the bulk.
%in anti-de Sitter space.
Then I argue that a coherent state 
in this string field theory
describes winding string condensation,
which has been
expected to cause 
the
topology change from the thermal AdS geometry
to
the AdS-Schwarzschild black hole geometry.
\end{abstract}

\end{center}

\setcounter{footnote}{0}
\end{titlepage}

\newpage
%%%%%%%%%%%%%%%%%%%%%%%%%%%%%%%%%%%%%%%%%%%%%%%%%%%%%%%
It has recently
been shown that the
low energy dynamics
of 
the large $N$
gauge theory on $S^3$ at finite temperature
reduces to a constant one-matrix model
\cite{Aharony:2003sx,Aharony:2005bq,Alvarez-Gaume:2005fv},
where the matrix is given by
the holonomy around the Euclidean time
direction compactified on a circle
(which I will call thermal circle).
One of the reasons that
the study in this direction has acquired interests 
is that via the AdS-CFT correspondence
\cite{Maldacena:1997re}
it is expected to describe
thermodynamics in anti-de Sitter space
\cite{Hawking:1982dh,Witten:1998qj,Witten:1998zw}.
There, at certain temperature
a phase transition
from the thermal AdS geometry
to
the AdS-Schwarzschild black hole
geometry takes place.
The transition
was identified with the
confinement-deconfinement transition
in the gauge theory side 
in \cite{Witten:1998qj,Witten:1998zw}.
One argument for this identification
is based 
on the calculation of the
Polyakov loop expectation value
from the bulk side.
It is given by the exponential of the 
(regularized) area of the
minimal surface in the bulk space-time
which ends on the Polyakov loop at the boundary
\cite{Rey:1998ik,Maldacena:1998im}.
Due to the difference of the topology
between the thermal AdS geometry
and
the AdS-Schwarzschild black hole geometry,
the Polyakov loops cannot have
expectation value in the former 
but can have in the latter.
It has further been suggested by many authors that
the change of the topology of the space-time
is caused by the tachyonic closed strings
winding around the thermal circle
\cite{KalyanaRama:1998cb,Barbon:2001di,%
Barbon:2002nw,Barbon:2004dd,McGreevy:2005ci,%
Horowitz:2006mr}.
To describe such condensation, it is desirable
to have a formalism that describes
quantum field theory of strings.
Actually,
there exists a prescription 
for constructing
a closed string field theory
in the temporal gauge\footnote{%
This parameterization of the worldsheet
time was first introduced in \cite{Kawai:1993cj}.} 
staring
from a given one-matrix model
\cite{Ishibashi:1993pc,Jevicki:1993rr,Ikehara:1994vx,%
Ikehara:1994mi,Ikehara:1995dd}
(see \cite{Ikehara:1994xs} for a review).
It is natural to identify
this 
as an effective closed string
field theory 
which
describes the
propagation of closed strings
in AdS in the radial and Euclidean time directions.
The Polyakov loops
are directly identified with
the closed strings
winding around the Euclidean time direction
in this formalism.

In this article,
I study the string filed theory in the temporal gauge
constructed from the matrix model 
obtained from the large $N$ 
gauge theory on $S^3$.
I will argue that
a vacuum state 
in this closed string field theory
can be consistently
interpreted as describing
the winding closed string (non-)condensation
in AdS space.
The closed sting field theory in 
the temporal gauge
has been used earlier 
in \cite{Periwal:2000dq}
to construct a simple model that
exhibits 
the AdS-CFT type 
correspondence via loop equations
\cite{Polyakov:1997tj,Polyakov:1998ju}.

The action of the
effective matrix model
for the Polyakov loops
has a form
\cite{Aharony:2003sx,Aharony:2005bq,Alvarez-Gaume:2005fv}
\bea
 \label{Seff}
S_{eff}(U,U^\dagger)
=
N^2\,
\sum_{k=2}^\infty\,
\sum_{n_1+n_2+\cdots+n_k= m_{k,{\bf n}}N}
c_{k,{\bf n}}\,
\rho_{n_1}\rho_{n_2}\cdots\rho_{n_k},
\eea
where ${\bf n}=(n_1,n_2,\cdots, n_k)$, 
$m_{k,{\bf n}}$ is some integer
and
\bea
 \label{rho}
\rho_n \equiv \frac{1}{N}\tr U^n ,
\eea
and $U$ is a $SU(N)$ unitary matrix.
In terms of the gauge theory 
on $S^3$ from which
one obtains the effective matrix model,
$\rho_n$ is nothing but the 
Polyakov loop winding around the
thermal circle for $n$ times:
\bea
\rho_n 
= 
\frac{1}{N}\tr P e^{i \int_0^{n\beta}dx^0 A_0},
\eea
where 
$x^0$ is the Euclidean time direction,
$\beta$ is the inverse temperature,
and $A_0$ is the 
time component of the gauge field.
$P$ denotes the path ordering.
The condition $n_1+n_2+\cdots+n_k=m_{k,{\bf n}}N$
is due to the $Z_N$ symmetry
$U \rightarrow U e^{2\pi i\frac{\ell}{N}}$,
$\ell = 1, \cdots, N-1$ which
the original gauge theory on $S^3$ had
(I will concentrate on such theories in this article).
The coefficients
$c_{k,{\bf n}}$ are in principle
determined by
the gauge theory on $S^3$ one considers
and depend on the temperature,
and $S_{eff}$ is real.
The matrix model
is given by
\bea
 \label{mm}
\int dU e^{-S_{eff}(U,U^{\dagger})}  ,
\eea
where
$dU$ is the left- and right- invariant 
Haar measure of the $SU(N)$ gauge group.

The loop equations, or the
Schwinger-Dyson equations,
in the leading order in $1/N$ expansion
are obtained as follows.
From the right translational invariance
of the Haar measure, one obtains
\bea
 \label{SD}
\left.
\int dU 
\frac{\pa}{\pa H_a}
\left\{
e^{-S_{eff}(Ue^{H},e^{-H}U^\dagger)} 
\left(
\frac{\pa}{\pa H_a}
\tr(Ue^H)^M
\right)
\right\}
\right|_{H=0}
=0,
\quad (M= \pm 1, \pm2, \cdots) 
\eea
where $H\equiv \sum_a H_a T_a$
is an arbitrary anti-Hermitian matrix
and $T_a$'s are 
normalized
anti-Hermitian
matrices which span the $N^2$ basis
of $N\times N$ matrices.\footnote{The convention here is 
$\tr T_a T_b = - \delta_{ab}$.} 
Since in this article I am mainly interested
in the $N \rightarrow \infty$ limit,
I will neglect the subleading differences
between $SU(N)$ and $U(N)$ gauge groups
to avoid
inessential complications. 
It is easy to incorporate them
in the following arguments.
By the identities for the $U(N)$ gauge group
\bea
\sum_a \tr A T_a B T_a = - \tr A \tr B
,\quad
\sum_a \tr A T_a \tr B T_a = - \tr AB,
\eea
one obtains polynomial equations
(loop equations):
\bea
 \label{loop}
f_M(z_n) =0, \quad (M= \pm 1, \pm2, \cdots) 
\eea
where 
\bea
z_n 
\equiv \langle \rho_n \rangle
\equiv 
\left \langle
\frac{1}{N} \tr U^n
\right \rangle
\equiv 
{\cal N}'
\int dU \, \frac{1}{N} 
\tr  U^n
\, e^{-S_{eff}(U,U^{\dagger})}
\eea
is the expectation value of the Polyakov loop
winding $n$-times around the thermal circle.
${\cal N}'$ is the normalization factor
${{\cal N}'}^{-1}=\int dU e^{-S_{eff}(U,U^{\dagger})}$,
and hence $z_0 =1$.
For example, in the simplest case
\bea
S_{eff} = c_2 \, \tr U \tr U^{\dagger},
\eea
one obtains
\bea
f_M(z_n)
=
M
\left(
c_2 \left( z_{M+1} z_{-1} -  z_{1} z_{M-1} \right)
+
\sum_{p=1}^M z_p z_{M-p} 
\right),
\eea
for $M \geq 1$, and similar equations
for $M \leq 1$.
Because of the $N\rightarrow \infty$ limit of
the $Z_N$ symmetry 
the matrix model action (\ref{Seff}),
the total winding number of the
each term appearing
in $f_M(z_n)$ is $M$.

From the 
loop equations eq.(\ref{loop}),
the tree level Hamiltonian $\hat{H}_0$
of the
closed string field theory 
in the temporal
gauge is given as follows:\footnote{%
The relative coefficients of 
$f_M$'s are determined 
so that they match with those coming from the
Fokker-Planck Hamiltonian in the stochastic
quantization, via the equivalence
between the temporal gauge quantization
and the stochastic quantization 
\cite{Jevicki:1993rr,Ikehara:1994xs}.
See the appendix for a brief summary
of the stochastic quantization and
the temporal gauge quantization.}
\bea
\label{TH}
\hat{H}_0
=
-
\sum_{\stackrel{M= -\infty}{M\ne 0}}^{\infty}
f_M(a_n^\dagger)a_M ,
\eea
where $a_n^{\dagger}$ and $a_n$
are the creation and annihilation operators
for the Polyakov loop with winding number $n$,
which I identify with the closed string
with winding number $n$.
They satisfy the usual harmonic oscillator
commutation relation
\bea
[a_n,a_m^{\dagger}] = \delta_{nm} \quad (n,m\ne0).
\eea
I assign winding number $n$ to the creation operator
$a_n^\dagger$ and $-n$ to the annihilation operator
$a_n$.
The $N\rightarrow \infty$ limit of
the $Z_N$ symmetry in 
the matrix model action (\ref{Seff})
leads to the conservation law
for the total winding number
during the time evolution
by the Hamiltonian (\ref{TH}).
Typically, 
there is a propagator term, and 
the tadpole term is forbidden,
as opposed to the case
studied e.g. in \cite{Ishibashi:1993pc}.
Notice that the closed strings 
do not merge during the time evolution
by the tree level Hamiltonian (\ref{TH}).

The first aim of the closed string field theory
in the temporal gauge is to
reproduce the 
expectation values of the Polyakov loops
in the matrix model as closed string
amplitudes:
\bea
\label{amp}
\lim_{t\rightarrow \infty}
\langle \Phi_f |
e^{- t \hat{H}_0}
a^{\dagger}_{n_1} a^{\dagger}_{n_2}
\cdots 
a^{\dagger}_{n_\ell}
|\Phi_i \rangle
=
\int dU e^{-S_{eff}}
\rho_{n_1} \rho_{n_2}
\cdots 
\rho_{n_\ell},
\eea
where $|\Phi_i\rangle$ 
is an initial ``vacuum" 
and $\langle \Phi_f |$ is a final 
tate %``vacuum" 
in the closed 
string field theory.
I will discuss more about them shortly.
In the planner limit
I am considering here, 
the amplitudes
should satisfy 
the factorization property
of the planer matrix model:
\bea
\label{tamp}
\lim_{t\rightarrow \infty}
\langle \Phi_f |
e^{- t \hat{H}_0}
g(a_n^{\dagger})
|\Phi_i \rangle
=
g(z_n),
\eea
where $g(z_n)$
is an arbitrary polynomial in $z_n$.

It is often the case that the choice
$| \Phi_i \rangle= | 0 \rangle$,
$\langle \Phi_f| = \langle 0 |$
in eq.(\ref{amp}) is appropriate.
Here, $|0\rangle$
is the usual harmonic
oscillator vacuum satisfying $a_n |0\rangle =0$.
The bra vacuum $\langle 0 |$
and ket vacuum $|0\rangle$ are assigned zero winding number.
The above choice however 
is not appropriate
for describing
a vacuum where the Polyakov loops
have expectation values, as I will explain
below.

Before going into the formalism,
I would like to comment on a
physical view in mind
based on the AdS-CFT correspondence:
I expect that the
time $t$ in the temporal gauge quantization
is related to the 
holographic
radial direction in the bulk.
Here, $t=0$ is interpreted 
as the asymptotic boundary
and 
$t \rightarrow \infty$ is interpreted
as deepest inside the bulk.

With the above view in mind,
I set the initial vacuum as
$|\Phi_i\rangle = |0\rangle$
which describes the state
with no closed string. 
The amplitude
(\ref{amp}) is expected to have a well-defined
$t\rightarrow \infty$ limit.
Then by taking $g(z_n) = z_M$ in (\ref{damp}) 
and differentiating by $t$,
one obtains
\bea
 \label{damp}
\lim_{t\rightarrow \infty}
\langle \Phi_f |
e^{-t\hat{H}_0}
f_M(a_n^{\dagger})
| 0 \rangle
= 0.
\eea
Let us take the (overcomplete)
coherent state basis:
\bea
 \label{coherentb}
\langle \Phi_f |
=
\langle \Phi(w_n) |
\equiv
\prod_{\stackrel{n = -\infty}{n\ne 0}}^{\infty}
\langle w_n|  ,
\eea
where 
$\langle w_n|$ is the coherent
state, i.e. the eigen-state of the
creation operator $a_n^{\dagger}$
with the eigen-value $w_n$:
$
 \label{coherent}
\langle w_n| a_n^{\dagger} 
= 
\langle w_n | w_n
$.  
It can be expressed
as $\langle w_n | = \langle 0 | e^{-{w_n}a_n}$.
Plugging this into eq.(\ref{damp}),
one obtains the loop equation (\ref{loop}).
Therefore, 
the stable final state 
which is invariant under the time translation
(this is the
relevant part of the final state
in the $t\rightarrow \infty$ limit)
is given by
$\langle \Phi_f|=\langle \Phi(z_n^c)|$,
where
the set of the eigen-values $\{ z_n^c \}$ 
satisfy the loop equations 
(\ref{loop}).\footnote{The loop equation
can have multiple solutions
in the case at hand.
Each solution corresponds
to a closed string field theory 
expanded around each classical saddlepoint.
Note that 
I haven't included non-perturbative effects 
in $1/N$ yet.
It will be briefly discussed
at the end of this article.
Also see the appendix.}
With this choice
the closed string field theory amplitudes
reproduce the matrix model results,
i.e. (\ref{tamp}) has been achieved.
Thus, whatever the physical motivations
I have given to set up the formalism,
the resultant 
formalism is justified
by that the first aim of 
the construction of 
the closed string field theory 
has been achieved.
There's no fundamental difficulty
in generalizing
this construction to the higher order
in the closed string loop expansion.

$\rho_n$ defined in eq.(\ref{rho}) 
is actually 
the $n$-th Fourier component
of the density distribution $\rho(\theta)$
of the eigen-values $A_{0a}$
of the zero-mode of the temporal component of the
gauge field:
\bea
\rho(\theta) 
\equiv
\frac{1}{N}
\sum_{a=1}^{N}
\delta(\theta - \beta A_{0a})
=
\frac{1}{2\pi}
\sum_{n=-\infty}^{\infty}
\rho_n e^{2\pi i n \theta}.
\eea
$A_0$ is related with 
the unitary matrix $U$ by
$U = e^{i \beta A_0}$.
As a density distribution,
$\rho(\theta)$ must satisfy
$\rho(\theta) \geq 0$, 
and it is normalized as 
$\int_{-\pi}^\pi\rho(\theta)=1$.
This also constrains the value of
$\rho_n$ as well as $z_n^c$.
Also, if the matrix model action
is real, one has
$z_{-n}^c = (z_n^c)^*$.

The confined phase in the gauge theory
corresponds
to $z_n^c =0$ ($n \ne 0$), 
i.e. the case where
all the expectation values
of the Polyakov loops vanish.\footnote{%
For a class of models,
$z_n^c = 0$ is always a solution
for the large $N$ saddlepoint equation of the
effective matrix model,
although it ceases to be the most dominant
saddlepoint after the deconfinement
phase transition.}
In this case $\langle \Phi(z_n)|$
is nothing but 
the harmonic oscillator vacuum $\langle 0 |$.
The harmonic oscillator vacuum
$\langle 0 |$ is naturally identified with 
the thermal AdS geometry in the bulk 
in the current formalism, 
for the reason described below.
This is
consistent with the expectations from
the known results \cite{Witten:1998qj,Witten:1998zw}.
The 
closed string states with non-zero winding 
number  
cannot dissappear 
into the final oscillator vacuum $\langle 0 |$
due to the 
winding number conservation
originating from 
the $N\rightarrow \infty$ limit of
the
$Z_N$ symmetry of the
matrix model.
This is interpreted in the bulk as follows:
In the thermal AdS topology,
due to the non-contractible
cycle around the thermal circle
the total winding number is conserved.
Thus 
the closed string with non-zero winding number
cannot dissappear into the vacuum in the bulk.
%since the vacuum has zero winding number.

In the deconfined phase of the gauge theory,
the Polyakov loops have non-zero expectation
values $z_n^c$.\footnote{At each of the
saddlepoints
related by the $Z_N$ symmetry.}
If looked from the harmonic oscillator vacuum
$\langle 0 |$,
$\langle \Phi(z_n^c) |$
with $z_n^c \ne 0$
is a state with winding
closed string condensation.
Since $\langle \Phi(z_n^c) |$
is a superposition
of the states with non-zero winding numbers,
the closed string states with non-zero number
can have overlap with it.
Thus the closed string tree amplitudes
with non-zero winding number in the initial
state and the vacuum at the final state
can be non-zero.
On the other hand,
regarding $\langle \Phi (z_n^c)|$ itself as
a new vacuum may correspond to
absorbing  
the effect of the
condensation into the space-time geometry.
In this interpretation,
there is no more string condensation
because it has already absorbed
into the geometry.
Then, the topology of the new space-time
should not have a non-contractible circle
so that the closed strings with non-zero winding 
number in the initial state
can dissappear in the final state in the bulk.
This is consistent with the expectation
from the known results
that the deconfined phase should correspond to the
AdS-Schwarzschild black hole geometry
which does not have non-contractible circle.

Thus, the final state
$\langle \Phi(z_n^c) |$ seems
to capture the difference of topology
in
the thermal AdS geometry and
the AdS-Schwarzschild black hole geometry.
The identification of $t\rightarrow \infty$
direction 
with the direction towards inside the bulk
nicely fit the UV-IR relation
in the AdS-CFT correspondence,
since the vacuum, 
or the phase of the boundary theory, 
is governed by the IR physics.
Since 
the vacuum expectation values 
of the Polyakov loops $z_n^c$
depend on the phase of the gauge theory,
$\langle \Phi (z_n^c)|$ is different
in different phases.
That the different space-time
can be described 
as different states
in the same theory
is an
realization of the
background independence.\footnote{%
See \cite{Ishibashi:1995in,Ishibashi:1996er} for a study
on this issue in the string field theory
in the temporal gauge.}
At the level of the effective
action (\ref{Seff}),
the change of the
temperature does look as a deformation of the theory,
since the coefficients $c_{k,{\bf n}}$
depend on the temperature.
However, if one looks at it
from the original gauge theory 
on $S^3$, this is only the change 
in the temperature,
the theory remains the same.

Finally, let us turn to
the closed string field theory
at finite $g_s=1/N$.
For this
one should consider the following
Schwinger-Dyson equation:
%instead of (\ref{SD}):
\bea
\label{SD2}
\left.
\int dU \frac{\pa}{\pa H_a}
\left\{
e^{-S_{eff}(Ue^{H},e^{-H}U^{\dagger})}
 \left(
  \frac{\pa}{\pa H_a}\tr (Ue^H)^{M_1}
 \right)
\tr(Ue^H)^{M_2}
\right\}
\right|_{H=0}
= 0.
\eea
Following the similar steps as
in the tree level case, one obtains
the full Hamiltonian $\hat{H}$:
\bea
\hat{H} = %\frac{1}{g_s}
\hat{H}_0 + g_s^2 \hat{H}_1,
\eea
where $g_s = {1}/{N}$,
and $\hat{H}_1$ has a form
\bea
\hat{H}_1 = 
\sum_{M_1,M_2 \ne 0}
M_1 M_2\,
a_{M_1+M_2}^\dagger
a_{M_1} a_{M_2}.
\eea
The term $\hat{H}_1$ in the Hamiltonian
introduces the merging of closed strings
during the time evolution.
If one considers at finite $1/N$,
the saddlepoint values of the
Polyakov loops are no longer exact.
By the similar reasonings
in the tree level case,
the 
%appropriate ground state 
final state stable under the time evolution
would be written as
\bea
 \label{sposition}
\langle \Phi_f | =
{\cal N}'\int dU e^{-S_{eff}}
\prod_{\stackrel{n=-\infty}{n\ne 0}}^{n=\infty}
\langle \rho_n | ,
\eea
if one allows to use
the answer from the
matrix model.
This is because again in
the coherent state basis
the existence of
the $t\rightarrow \infty$
limit implies the loop equations,
and the expectation values of
the Polyakov loops should satisfy those.
Once the 
non-perturbative effects in $1/N$
are taken into account
as in (\ref{sposition}),
the ground state
no longer allows a 
simple geometrical interpretation
of the bulk space-time,
but rather it gives some sort of 
quantum superposition of geometries
\cite{Maldacena:2001kr}.
I leave further study of those $1/N$
effects to the future investigation.

To summarize,
in this article
I argued that the tree level
vacuum $\langle \Phi(z_n^c)|$
in the closed string field theory
in the temporal gauge
can be consistently interpreted
as describing the
winding string (non-)condensation
in AdS apace at finite temperature.
The closed string field theory
was constructed so that it satisfies 
the loop equations
of the matrix model of the Polyakov loops.
The condensation of the winding strings
can also be interpreted as a change
of the space-time topology in the bulk.
The 
Polyakov loop expectation values
in the matrix model was directly
translated into the 
winding string condensation
in this formalism.
It is nice that the winding string condensation
was described in the quantum field theory
of closed strings.
It is also nice that
the temporal gauge quantization gives
a way to understand
how in the loop equations
the
differences of the phases in the gauge theory
side is reflected to the differences in the 
bulk geometry.

Critical readers may think that
the closed string field theory
constructed above
can describe 
at best what the matrix model can describe,
and provides no new information.
But in my view,
the AdS-CFT correspondence
looks more well defined in
the CFT side, at least in the 
weak 't Hooft coupling region.
In most cases
we do not really have a satisfactory formulation
of quantum field theory of closed strings, and 
not even sure whether such formalism exists.
And the bulk space-time seems to be an
emerging concept arising
from the large $N$ limit.
It is possible that the
AdS-CFT correspondence
is not really a 
duality but rather a definition
of the bulk theory by
the boundary field theory.\footnote{%
See \cite{Hanada:2004im} for a recent discussion
on this issue in a related context.}
From this viewpoint,
it is important
to give descriptions to
the concepts in the
bulk in terms of the gauge theory language.
I hope the result of this article 
is a nice first step
for the direct 
translation of 
the expectation
values 
of the Polyakov loops in the 
gauge theory side 
into
the winding string condensation 
in the
closed string field theory language.

I also hope this study has shed some light
on the
background independence issue in
the closed string field theory
in the temporal gauge.

It seemed natural to
identify
the time
in the temporal gauge quantization
and the holographic radial direction
in the bulk space-time.
This point should be investigated further.\footnote{%
For earlier studies in this direction, see
\cite{Periwal:2000dq,Lifschytz:1999fw,Lifschytz:2000bj}.}
In particular,
relation to the minimal surface calculation
in the bulk geometry 
should be clarified.
It will also be interesting
to apply the temporal gauge quantization
to the full four-dimensional gauge theory.

The $1/N$ corrections should be also studied,
as well as
the cases where the double scaling limit
is relevant and 
higher closed loop 
contributions appear even at $N \rightarrow \infty$ limit, 
like those studied in \cite{Alvarez-Gaume:2006jg}.

\vspace*{3mm}
\begin{center}
{\bf Acknowledgments}
\vspace*{1mm}\\
I am very thankful to the people in India
for the research I could do here.
\end{center}

%%%%%%%%%%%%%%%%%%%%%%%%%%%%%%%%%%%%%%%%%%%%%%%%%%%%%%%%%
\appendix
\section*{Appendix}

In this appendix, 
I briefly summarize 
the stochastic quantization and
the temporal gauge quantization.
The explanations
on the temporal gauge quantization
might not be standard.
It is arranged so that
it becomes suitable for the purpose of
the current article.
The discussions on the multiple saddle points
at the end might not have been explored much before.
One can find standard reviews in
e.g. \cite{Ikehara:1994xs} and
\cite{Damgaard:1987rr}.

To illustrait the main points,
I will take a simple example:
A scalar field $\phi$ in zero-dimension.
The application to more general cases
would be obvious once the main idea is
understood.
What is of interest 
is the expectation value 
expressed by the path integral
\bea
\langle
F(\phi)
\rangle
=
\frac{%
\int d\phi \, e^{-S(\phi)} F(\phi)}{%
\int d\phi \, e^{-S(\phi)}} ,
\eea
where $S(\phi)$ is the action and
$F(\phi)$ is an arbitrary function of $\phi$.
The essential points of relevance in 
the stochastic quantization
is to consider the following
Fokker-Planck equation:
\bea
 \label{FPeq}
\frac{\pa}{\pa t} P(t;\phi)
=
-\frac{\pa}{\pa \phi}
\left(
\frac{\pa}{\pa \phi} +\frac{\pa S}{\pa \phi} 
\right) P(t;\phi)
\equiv
-H_{FP}^\dagger P(t;\phi).
\eea
The probability distribution
$P(t;\phi)$ satisfies
\bea
\int d\phi \, P(t;\phi) = 1.
\eea
The formal solution of the Fokker-Planck equation
(\ref{FPeq}) is given by
\bea
 \label{formal}
P(t;\phi) = e^{-t H_{FP}^\dagger} 
P(0;\phi).
\eea
From (\ref{formal}), one obtains
\bea
\int d\phi\, F(\phi) P(t;\phi)
=
\int d\phi\, F(\phi) e^{-t H_{FP}^\dagger} 
P(0;\phi) 
=
\int d\phi\, 
\left(e^{-t H_{FP}} F(\phi)\right) 
P(0;\phi) ,
\eea
where
\bea
 \label{tHam}
H_{FP} \equiv
-
\left(
\frac{\pa}{\pa \phi} - \frac{\pa S}{\pa \phi} 
\right)
\frac{\pa}{\pa \phi} .
\eea
I have assumed that there were no boundary terms
in the partial integration,
which requires $e^{-S(\phi)}$ to decrease
at $\phi \rightarrow \infty$ fast enough.
Note that the differential 
operator $H_{FP}$ is essentially the one
appeared in
the Schwinger-Dyson equations.
The Fokker-Planck Hamiltonian $H_{FP}$
can be made Hermitian by the
similarity transformation:
\bea
e^{-\frac{1}{2}S(\phi)}H_{FP}e^{\frac{1}{2}S(\phi)}
=
-
\left(
\frac{\pa}{\pa \phi} - \frac{1}{2}\frac{\pa S}{\pa \phi} 
\right)
\left(
\frac{\pa}{\pa \phi} + \frac{1}{2}\frac{\pa S}{\pa \phi} 
\right)
\eea
which manifestly has
positive semi-definite eigen-values.
After
taking the $t \rightarrow \infty$ limit,
one is expected to get the 
%equilibrium 
stationary configuration:
\bea
 \label{limit}
\lim_{t\rightarrow \infty} P(t;\phi) = e^{-S(\phi)}.
\eea
Thus 
\bea
\int d\phi\,
\lim_{t\rightarrow \infty} P(t;\phi) 
F(\phi) 
= 
\int d\phi \,
e^{-S(\phi)}
F(\phi) ,
\eea
which is what we wanted to get.

The temporal gauge quantization
can be interpreted\footnote{%
See 
\cite{Ikehara:1994xs} 
and references therein
for 
the original construction motivated from a
matrix model description of 
the discretized worldsheet.}
as the equivalent rewriting
of the above formula
using the
creation and annihilation operators
$\hat{\phi}^{\dagger}$, $\hat{\phi}$
which 
satisfy
\bea
[\hat{\phi},\hat{\phi}^{\dagger}]=1.
\eea
One interprete the previous formulas as
the 
``position space
representation":\footnote{%
In the main text I have used the
coherent states which were
more natural for the complex variables.}
\bea
\phi \leftrightarrow \hat{\phi}^{\dagger}, &&
\frac{\pa}{\pa \phi}  \leftrightarrow \hat{\phi},\nn
f(\phi) = \langle \phi| f(\hphi^{\dagger})|\hphi =0\rangle,&&
\langle \phi | \hat{\phi}^{\dagger}
=
\langle \phi | \phi.
\eea
Then, one can ``go back"
from the position space representation
to the operator expression:
\bea
\int d\phi \, 
\left(
e^{-t H_{FP}(\phi,\pa/\pa\phi)}
F(\phi)
\right) 
P(0;\phi)
=
\int d\phi \,
\langle \phi | e^{-t \hat{H}_{FP}(\hphi^\dagger,\hphi)}
F(\hat{\phi}^{\dagger})
|\hphi =0\rangle 
P(0;\phi). 
\eea
where $\hat{H}_{FP}$ is identified with
the Hamiltonian
in the temporal gauge quantization.
The final state of the
temporal gauge quantization
is defined as
\bea
 \label{fstate}
\langle \phi_f |
\equiv
\int d\phi \,
P(0;\phi)
\langle \phi | ,
\eea
so that one obtains
the equivalence between
the temporal gauge quantization
and the stochastic quantization:
\bea
 \label{equivalence}
\lim_{t \rightarrow \infty}
\langle \phi_f | e^{-t\hat{H}_{FP}} 
F(\hphi^\dagger)|\hphi =0 \rangle
\int d\phi \, e^{-S(\phi)}F(\phi)
\eea
Since $F(\phi)$ is an arbitrary function,
it must follow that
\bea
 \label{final}
\lim_{t\rightarrow \infty} 
\langle \phi_f |
e^{-t\hat{H}_{FP}}
=
\int d\phi \,
e^{-S(\phi)} \langle \phi | .
\eea
The saddlepoint approximation is
applicable,
e.g. 
when recovering the Planck constant
$S(\phi) \rightarrow \frac{1}{\hbar}S(\phi)$
and taking $\hbar \rightarrow 0$ limit:
\bea
 \label{sd}
\lim_{t\rightarrow \infty} 
\langle \phi_f |
e^{-t\hat{H}_{FP}}
=
\sum_{saddle\, points}
e^{-\frac{1}{\hbar}S(\phi_0^i)}
\langle \phi_0^i |,
\eea
where $\phi_0^i$'s are the saddle points.
I took into account the possibility
of multiple saddle points.
Suppose
one chooses to have a sharp initial
distribution
$P(0;\phi)=\delta(\phi-\phi_0^i)$ 
on one of the saddle points.
In the $\hbar \rightarrow 0$ limit,
one also obtains
$\langle \phi_0^i| \hbar \hH_{FP} = 0$
(see eq.(\ref{tHam}) with $\hbar$ recovered).
Then
one cannot get sum over the saddle points
in (\ref{sd})
but only $e^{-S(\phi_0^i)}$ 
for that one saddle point.
%Differentiating both sides by $t$,
%In the $\hbar \rightarrow 0$ limit,
%one also obtains
%$\langle \phi_0^i| \hbar \hH_{FP} = 0$.
This was basically 
what happened in the main text
($1/N^2$ played the role of $\hbar$).
At the tree level,
the closed string field theory
was build on each saddle point.
In a sense, considering
each saddle point separately in this way
is an artifact of $\hbar \rightarrow 0$ limit
which leads to the limitation of the
stochastic quantization
with the sharp initial
distribution $P(0;\phi)=\delta(\phi-\phi_0)$.

When applying the stochastic quantization
or the
temporal gauge quantization
to the loop equations,
the change of variables
from the original matrix fields to loops
is involved.
See \cite{Sakita:1979gs,Jevicki:1979mb,%
Jevicki:1980zg,Jevicki:1993rr} on this aspect.

%%%%%%%%%%%%%%%%%%%%%%%%
\bibliography{tsft}
\bibliographystyle{kazu}
%%%%%%%%%%%%%%%%%%%%%%%%

\end{document}